# Comparison of Fluid Attenuated Inversion Recovery Sequence with Spin Echo $T_2$-Weighted MRI for Characterization of Brain Pathology

Indra Dev Sahu, [1, 4] Sheshkant Aryal, [2] Shanta Lal Shrestha, [3] Ram Kumar Ghimire [3]

[1]Current address, Department of Physics, University at Albany, State University of New York, 1400 Washington Avenue, Albany, NY, 12222, USA

[2]Central Department of Physics, Tribhuvan University, Kirtipur, Kathmandu, Nepal

[3]Institute of Medicine, Tribhuvan University, Teaching Hospital, Maharajgung, Kathmandu, Nepal

[4]Corresponding Author

## Abstract

Twenty cases of different brain pathology have been studied via MRI using an open resistive magnet with magnetic field strength of 0.2 Tesla. The relative signal intensity with respect to the repetition time (TR) at fixed echo time (TE) 0.117 sec. has been studied. It was found that the signal intensity saturates for most lesions beyond a certain TR~6 sec in the $T_2$ - weighted image. The signal intensity differs with respect to the inversion time (TI) for fat and cerebrospinal fluid (CSF). It was found that the intensity is nulled for CSF at TI ~1.5 sec. and for Fat at TI~0.10 sec in the FLAIR imaging sequence. Thus the intensity of the lesions is qualitatively different for the two sequences. From the radiological diagnostic point of view, it was concluded that the FLAIR sequence is more useful for the detection of lesions compared to $T_2$ sequences.

## Introduction

Magnetic resonance imaging (MRI) is a powerful, non-invasive medical technology that allows physicians to identify and study pathologies in biological tissues. The tissues examined can be represented as sectional images showing great contrast between healthy and stressed tissue with high resolution.

The MR signals from various tissue segments are displayed as video pixels. The intensity of a signal and the brightness of the pixels depend on the $T_1$ and $T_2$ values, concentration of hydrogen protons fluid flow. Fats appear very bright in MR images, the air cavities are dark. Fluid flow through the tissue being studied is primarily related to blood circulation which most often appears dark in the MR image. To improve the quality of clinical information that may be obtained from an MR image, one may use a variety of techniques. The most important ones are: 1) $T_1$ Weighted Imaging Technique 2) $T_2$ Weighted Imaging Technique 3) Fluid Attenuated Inversion Recovery (FLAIR) Imaging Technique 4) Short Time Inversion Recovery (STIR) Imaging Technique 5) Magnetic Resonance Pulse sequence Technique. Tissue signal suppression may be accomplished using STIR (short time inversion recovery) for systems with short $T_1$'s. A variation of STIR is the FLAIR technique (Fluid -attenuated inversion recovery), which helps to identify even very small demyelinating lesions (e. g., multiple sclerosis). [1]

The diagnostic accuracy of MR imaging is dependent on the selection of imaging sequence parameters that will enhance image quality. Combining RF pulses with gradients of the magnetic field allow one to resolve spatial information that results in a high-resolution, informative image. Since different brain tissues have different characteristic relaxation times, $T_2$-weighted imaging and FLAIR imaging can be used to resolve different tissue stresses. The $T_2$ image gives high intensity while FLAIR nulls the signal intensity of CSF. In some cases the FLAIR image can be even more intense. In order to delineate the brain lesions more clearly when displayed as a $T_2$-weighted image, both types of the sequences are used while performing

MRI. This study helps not only to quantify different brain lesions but also helps us to know which particular sequence is better for a given brain pathology.

## Materials and Methods

The present study was carried out at the department of Radiology and Imaging, TU Teaching Hospital Maharajgunj, Kathmandu, Nepal. The MRI unit was an open type 0.2 Tesla Magneton manufactured by Siemens AG Germany. A detail description of components in a modern MR system and the basic composition of MR imaging system can be found in references. [2, 3, 4]

Twenty cases of different brain pathology were studied using an open resistive magnet with a magnetic field strength of 0.2 Tesla. The signal intensity and characteristics have been compared in both the $T_2$-weighted MRI sequence and the FLAIR sequence. The relative signal intensity with respect to the repetition time (TR) at fixed echo time (TE) ~0.11 sec. was studied. The (TR/TE/TA) used in the FLAIR sequence is (6 sec/0.095 sec./7 min. 19 sec.) and in $T_2$-weighted sequence is (6 sec/0.117 sec./7 min. 19 sec.).

During the experiment, the 20 patients were studied using $T_1$-weighted sequence, spin echo $T_2$-weighted sequence and FLAIR sequence in different planes such as sagittal, transversal and coronal. The images of all the patients have been studied. In this report, we show a representative sample of the data set.

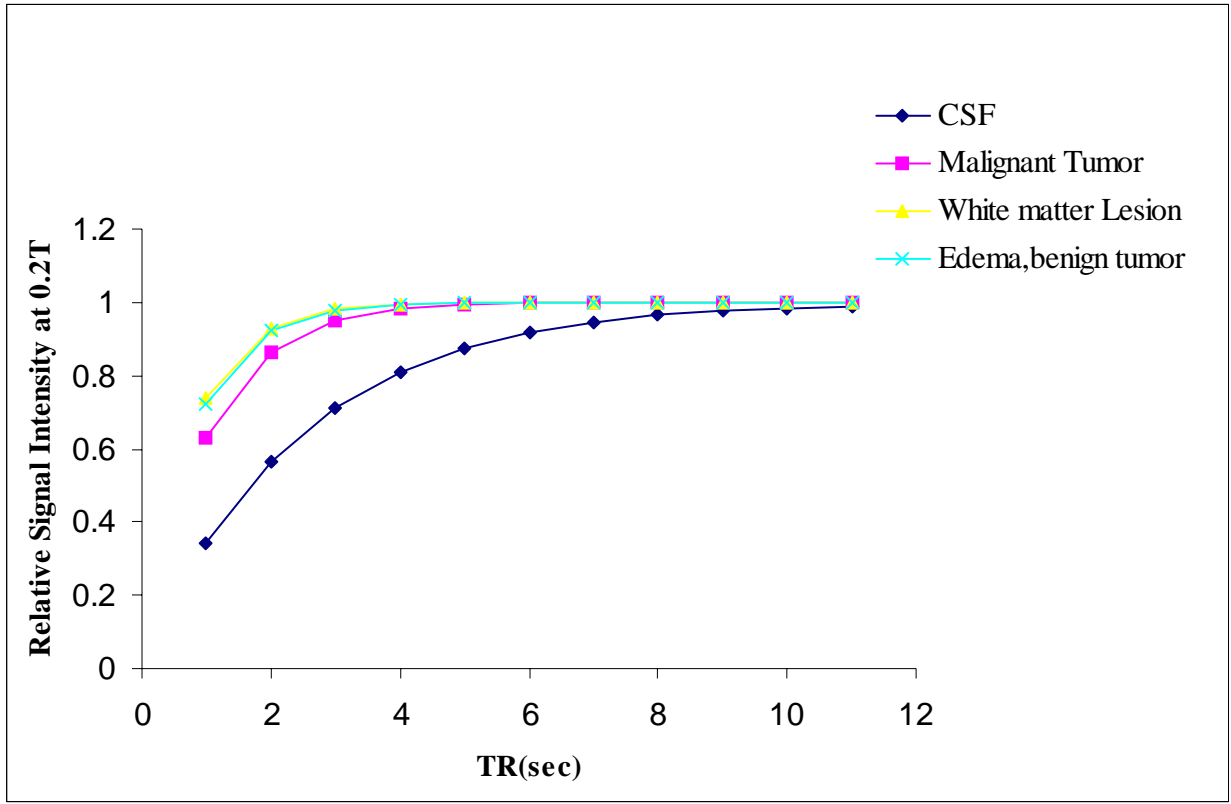

Figure 1(a) Relative signal intensity for CSF, White matter lesion, Edema, Benign tumor and Malignant tumor against the repetition time at fixed echo time (TE=0.117 sec ) for $T_2$-weighted sequence.

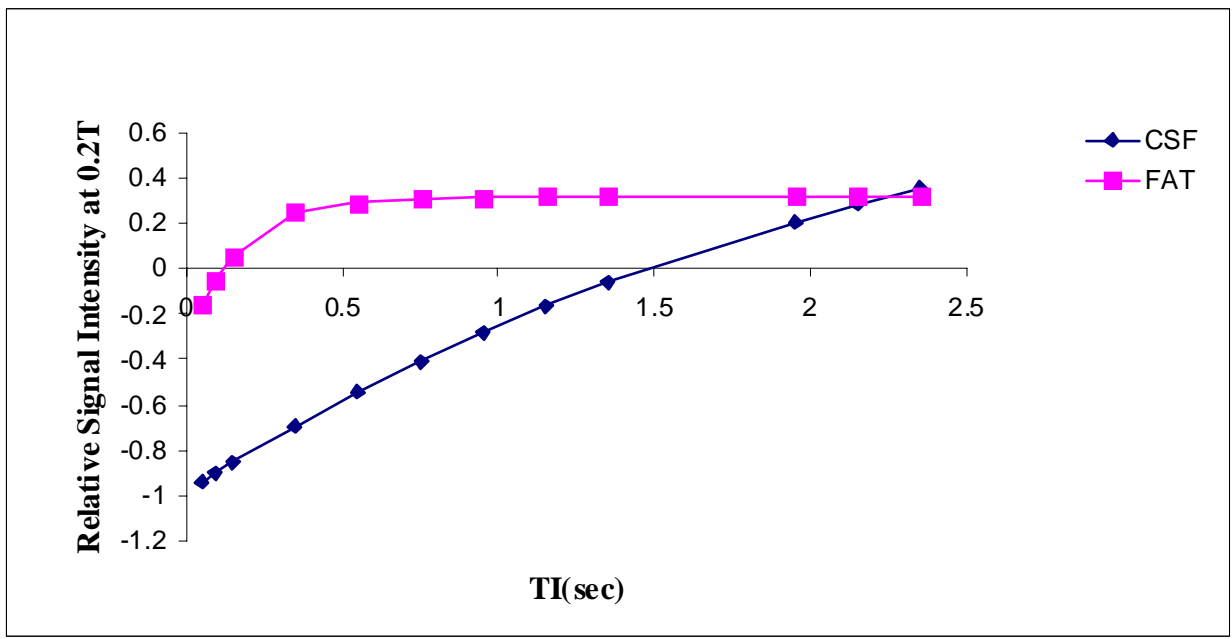

Figure 1(b) Relative signal intensity for CSF and fat at different inversion time for fixed TR=6000ms and echo time =95ms for FLAIR.

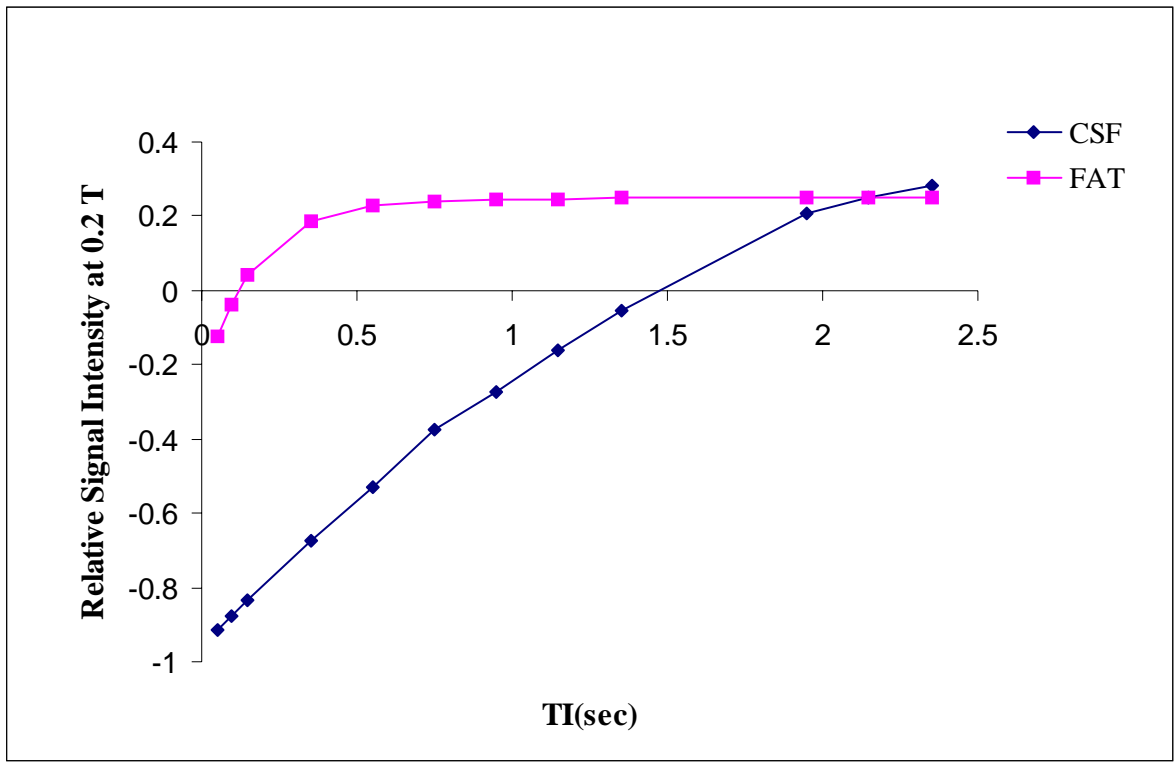

Figure 1(c) Relative signal intensity for CSF and fat at different inversion time for fixed TR=6000ms and echo time=117ms for FLAIR.

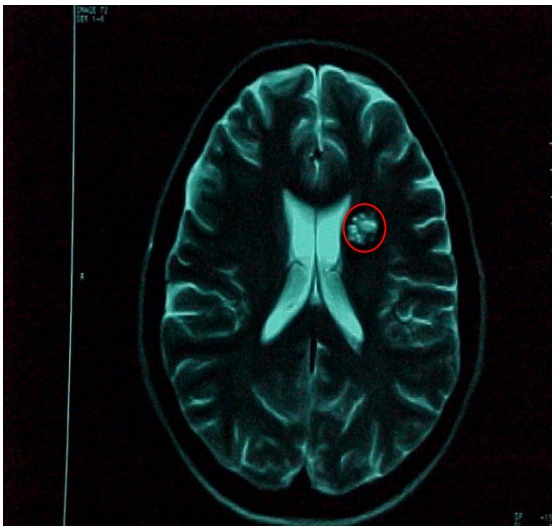

Figure 2 $T_2$- weighted axial sections showing high signal area in left basal ganglia.

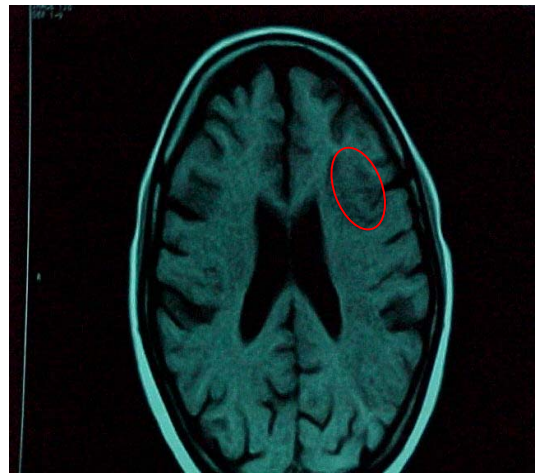

Figure 3 $T_1$-weighted image showing slightly low signal area in left temporal and parietal lobe.

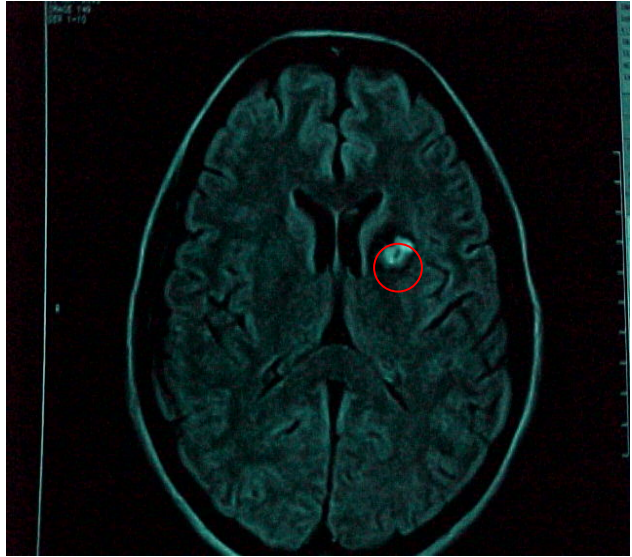

Figure 4 FLAIR axial section showing high signal intensity area in left basal ganglia.

## Results and Discussion

The relative signal intensity for CSF, White matter lesion, Edema, Benign tumor and Malignant tumor against the repetition time at fixed echo time (TE= 0.117 sec) in the case of the spin -echo $T_2$ weighted sequence is plotted as shown in Figure (1) . From the graph in Figure 1(a), it is clear that as TR increases, the intensity saturates. At low TR, the intensity is small. In order to obtain good signal intensity we typically used TR=6000ms. The relative signal intensity for CSF and fat at different inversion times were taken under the following conditions: TR = 6000ms and Echo time = 95ms. The details are shown in Figure 1(b) and 1(c). From the graph, the FLAIR - Sequence nulls the CSF signal at an inversion time of (~1500 ms). The fat signal is suppressed at (~100 ms). This sequence can therefore separate signal from brain lesions and

CSF. With an appropriate choice of inversion time, brain lesions can be displayed with higher resolution with FLAIR.

The images of all the patients have been studied. Some photographs have been shown in Figure (2), (3) and (4) along with medical description. On the basis of radiological diagnosis of the 20 patients, the following conclusions have been drawn.

1) In cases of the lymphoma /GBM (Glioblastoma Multiforme) lesion, intensity of the lesion signal is high in both the $T_2$ and FLAIR weighted sequence , but the periventricular , cortical lesion and cystic degeneration are more clearly identified in FLAIR .

2) For the Multiple Sclerosis/Ischaemic relfaction, Chronic infarcts and Neuro cysticercosis, the intensity is high in $T_2$ and low in FLAIR with high intensity peripheral rim separable from CSF in FLAIR. Cystic area, Perilesional edema and Margin are more clearly seen in FLAIR.

3) In case of vein of Galam aneurysm with hydrocephalus, AVM (Arterio Venous Malformation) and healed granuloma, the intensity is low but AVM is more clearly seen in $T_2$.

4) For pontine glioma, cerebellar infarction, the intensity is high with no strong reference for $T_2$- or FLAIR Spectroscopy. This means that these lesions can be detected by applying either of these sequences.

Okuda Tomoko et al. concluded qualitatively and quantitatively that the contrast of brain lesions detected with FLAIR images provided useful information. Except in the cases of Multiple sclerosis, either FLAIR or intermediate-weighted sequences should be added to $T_2$ weighted sequences for MR imaging.[5] Tsuchiya Kazuhiro et al. compared the unenhanced and contrast-enhanced FLAIR imaging with other sequences to visualize meningeal carcinomatosis and concluded that unenhanced FLAIR images are of more value than spin-echo $T_2$-weighted images for the diagnosis of intracranial meningeal carcinomatosis.[6] Hashemi Ray H. et al. compared thin-section, sagittal, fast FLAIR MRI with conventional axial spin-echo (SE) imaging for detection of multiple selerosis (MS) in the brain and found that sagittal thin- section fast FLAIR is superior to conventional axial proton-density and $T_2$-weighted SE pulse sequence for detection of MS plaques in the brain.[7] Mathews Vincent P. et al. determined the utility of gadolinium-enhanced FLAIR MRI of the brain by comparing results of gadolinium-enhanced $T_1$-weighted MR imaging with magnetization transfer (MT) saturation and found that fast FLAIR images have noticeable $T_1$ contrast making gadolinium induced enhancement visible.[8]

## Conclusion

$T_2$ weighted images and FLAIR can both provide useful diagnostic information in the study of brain lesions. In some cases the $T_2$ sequence may be preferable but in the majority of cases the FLAIR sequence is preferred. Although the noise level is higher in FLAIR images as compared to Spin Echo $T_2$ Weighted Imaging, FLAIR is more useful than the $T_2$ weighting sequence for the detection of brain pathology.

FLAIR sequences are thus useful for studying: infarction, multiple sclerosis (MS), subarachnoid hemorrhage, head injury and meningeal carcinomatosis without use of contrast agent. One drawback of FLAIR is that it requires a longer acquisition time than competing techniques, but its superior resolution makes it a valuable diagnostic tool.

## Acknowledgement

We are grateful to Central Department of Physics, Tribhuvan University, Kirtipur Kathmandu, Nepal and Institute of Medicine, Tribhuvan University, Kathmandu, Nepal, for providing their facilities during this research work. We are also grateful to the Department of Physics, SUNY at Albany, New York, USA